# Simple ΔV Approximation for Optimization of Debris-to-Debris Transfers

Hong-Xin Shen [1]
*Xi'an Satellite Control Center, 710043 Xi'an, China*

Lorenzo Casalino [2]
*Politecnico di Torino, 10129 Torino, Italy*

A method for the rapid estimation of transfer costs for the removal of debris in low Earth orbit is proposed. Debris objects among a population with similar inclination values are considered. The proposed approximate analysis can provide estimations of actual Δv between any debris object pair as a function of time; these estimations allow for the rapid evaluation of the costs of large sequences of targets to be removed. The effect of Earth's oblateness perturbation (J2) is exploited to reduce transfer costs. The debris removal problem of the 9$^{th}$ edition of the Global Trajectory Optimization Competition is used to evaluate the estimation accuracy; Δv estimations of the transfers between objects pairs are verified by comparing them with the GTOC9 solution proposed by the winning team from JPL. The comparison of the results demonstrates the very good accuracy of the simple approximation.

## I. Introduction

A large number of space debris occupies the Earth orbit regions of most interest, mainly in low-Earth orbit (LEO) and geostationary orbit (GEO). Operational satellites and manned space vehicles or stations are threatened to be destroyed by space debris. Kessler Syndrome [1] states that more and more debris might be produced due to collision events, even if all launches into space would be stopped immediately. Only remediation of the debris presence in the near-Earth environment and avoidance of future problems for research in and commercialization of space is the removal of existing large objects from orbit [2]. Thus, active debris removal is of great relevance. Recent studies have led to the conclusion that, in addition to mandatory end-of-life disposal of new satellites, removing about 5 objects per year may be necessary to stabilize the future LEO debris population [3].

Many LEO objects occupy orbits with similar altitude (but not exactly the same) at particular inclinations, that is, the most useful for space applications (e.g., Sun-synchronous, Molniya). Clustered objects, however, do not share the same orbit plane, even though they have the same inclination, as right ascension of ascending node (RAAN) is perturbed due to Earth oblateness. Its effect depends on the orbital elements (mainly, the semimajor axis and the inclination). For this reason, RAAN values are typically spread almost uniformly. By taking advantage of Earth oblateness perturbation, a few debris objects with approximately the same inclination could potentially be removed in LEO at little cost, using differential nodal precession rates.

---

[1] Assistant Professor, State Key Laboratory of Astronautic Dynamics; hongxin.shen@gmail.com.
[2] Associate Professor, Dipartimento di Ingegneria Meccanica e Aerospaziale, Corso Duca degli Abruzzi, 24; lorenzo.casalino@polito.it.

In the typical scenario, an active debris removal spacecraft performs rendezvous with the target and then provides it a velocity change, to place it on a reentry trajectory. There is an obvious advantage if the same ADR mission can remove more than a single object. In viable mission architectures, the chaser attaches a deorbiting module to the target and then moves towards the following object. Choosing the sequence of debris to be removed is crucial to the cost-effectiveness of such missions. The sequence optimization results in a combinatorial problem, which has been studied using approaches such as branch and bound algorithms [4], ant colony optimization [5], column generation techniques [6].

Because of the dimensionality curse of the combinatorial problem, it could be essential to analytically estimate the body-to-body transfer cost, in order to enable broad search in a reasonable computational time. Previous works generally estimated the transfer $\Delta v$ by adding or taking the root-sum-square of the individual $\Delta v$s needed to match the semi-major axis, node angle, and inclination of the target debris [7]. Alfriend et al. presented an analytical estimation for satellite-to-satellite transfer in order to get an optimal servicing of multiple targets, but the focus was on Geosynchronous orbits [8] where J2 has a negligible effect. When considering the RAAN match in LEO for a specified transfer time, the general strategy consists in waiting for RAAN drift to align the orbit plane. However, in many circumstances this strategy may not be pursued.

In this paper, a simple analytical method is developed to make sure that the transfer costs between two objects of the removal sequence can be accurately approximated. The accuracy of these estimations is then verified by comparison with existing extensively optimized solutions. Rendezvous transfers between two given orbits are dealt with using an accurate dynamical model that takes J2 perturbation into account. The state-of-the-art solution winner the ninth Global Trajectory Optimization Competition (GTOC9) by JPL [9] is used as benchmark to verify the accuracy of the method presented in this paper.

## II. Dynamic model

Space debris dynamics are usually defined with a set of Two-line Elements (TLE) and the SGP4 propagator [10]. However, the accuracy of debris ephemerids degrades with time, and the TLE information has to be updated regularly. Thus, a simplified propagation model is adopted here to describe the dynamics of the debris.

The most interesting targets for removal missions have relatively large values of semi-major axis; at the corresponding altitudes atmospheric drag is too low to slow down the debris remarkably, and is here neglected. The body orbit is described by means of osculating orbital elements, which are specified at the known ADR mission departure time. Only secular orbit perturbation due to Earth oblateness (related to harmonics term J2) is then considered. Semimajor axis $a$, eccentricity $e$ and inclination $i$ are constant, whereas right ascension of ascending node $\Omega$, argument of periapsis $\omega$ and mean anomaly $M$ vary according to

$$\frac{d\Omega}{dt} = -\frac{3}{2}\sqrt{\frac{\mu}{a^3}}\frac{J_2 \cos i}{(1-e^2)^2}\left(\frac{r_E}{a}\right)^2 \tag{1}$$

$$\frac{d\omega}{dt} = \frac{3}{4}\sqrt{\frac{\mu}{a^3}}\frac{J_2(5\cos^2 i - 1)}{(1-e^2)^2}\left(\frac{r_E}{a}\right)^2 \tag{2}$$

$$\frac{dM}{dt} = \sqrt{\frac{\mu}{a^3}} + \frac{3}{4}\sqrt{\frac{\mu}{a^3}}\frac{J_2(3\cos^2 i - 1)}{(1-e^2)^{3/2}}\left(\frac{r_E}{a}\right)^2 \tag{3}$$

where $\mu$ is Earth's gravitational parameter, $r_E$ is Earth's equatorial radius.

In case of low eccentricity, the deviation of RAAN rate ($\delta\dot{\Omega}$) due to small changes of $a$ ($\delta a$) and $i$ ($\delta i$) can be obtained by taking derivative of $\dot{\Omega}$ to $a$ and $i$

$$\frac{\delta\dot{\Omega}}{\dot{\Omega}} = -\frac{2}{7}\frac{\delta a}{a} - \tan i \delta i \tag{4}$$

Given starting values of orbital elements, the perturbed values can be evaluated at any time $t$, and position and velocity are consequently determined. This propagation model has been contrasted to SGP4 propagation to compare their accuracy [5]. Great accordance has been found even for 200-day propagation (errors below few kilometers) in the evaluation of orbit shape and orientation. Large errors are found for argument of perigee and mean anomaly. However, the effect of errors in $\omega$ is small due to the limited eccentricity of the orbits, and phase adjustments have a small impact on $\Delta v$, thanks to the relatively large number of revolutions during each transfer leg. Thus, the propagation accuracy is considered to be satisfactory.

## III. Approximate Transfer Cost Evaluation

The chaser spacecraft is inserted by the launcher into rendezvous conditions with one of the objects in the debris set (target 1), for its removal. The mission starts at initial time $t_1=0$ and the chaser orbital elements are those of the initial object at $t_1$. The chaser then moves to perform rendezvous the following target and so on, until the last object has been reached. The present work focuses on finding a way to estimate the transfer cost between objects pairs as a function of transfer time, in order to speed up the performance of combinatorial optimization algorithms that define the best object sequences.

In general, the transfer time between any debris pairs is estimated based on the assumption that favorable opportunities occur only when the required plane change is small, and therefore when the RAAN of the chaser is close to that of the target. To this purpose, the mission can take advantage of J2 perturbation, which changes the RAAN of bodies orbiting the Earth with a rate that depends on semi-major axis and inclination. Objects with different orbits will therefore have different rates of change of $\Omega$. However, cases when this kind of solutions are not suitable may occur, e.g., in the presence of strict time constraints, and a different estimation method is required.

### A. The transfer time is optimal

The transfer time from debris $k$ to $k+1$ can be easily evaluated. Optimal phasing is assumed to obtain an estimation of the transfer $\Delta v$. The Hohmann transfer cost for negligible radius change reduces to $\Delta v / v = 0.5 \Delta r / r$. The change of semimajor axis $\Delta a$ is here considered and an empirical relation is introduced to account for the additional eccentricity vector change $\Delta e$ [11]

$$\Delta v / v = 0.5 \sqrt{(\Delta a / a)^2 + \Delta i^2 + \Delta e^2} \qquad (5)$$

The average value of semi-major axis $a$ of the two objects involved in the leg, and the corresponding circular velocity $v$ may be used.

According to the above analysis, the theoretical trip time and velocity change between any debris pair can be determined. It is worth noting that a change of arrival time for leg $k$ usually does not propagate to the following legs, as convenient transfers typically require relatively long waiting times before starting to maneuver. This approximation has been validated with good accuracy, as shown in a previous work [5].

**B. The transfer time is limited**

The transfer time from debris $k$ to $k+1$ can be imposed as $t$. Eq. (1) provides the final difference of RAAN between the chaser and the target. It is assumed that $\Omega_{k+1}(t) - \Omega_k(t) \neq 0$, $a_{k+1} - a_k \neq 0$, $i_{k+1} - i_k \neq 0$; the corresponding maneuvers need velocity changes denoted by $x$, $y$, $z$ [11][12]:

$$x = (\Omega_{k+1}(t) - \Omega_k(t)) \sin i_0 v_0 \quad (6)$$

$$y = \frac{a_{k+1} - a_k}{2a_0} v_0 \quad (7)$$

$$z = (i_{k+1} - i_k) v_0 \quad (8)$$

where $a_0 = (a_{k+1} + a_k)/2$, $i_0 = (i_{k+1} + i_k)/2$, and $v_0 = \sqrt{\mu/a_0}$.

A two-impulse transfer is considered in this approximate analysis. Without loss of generality, the first impulse completes a partial change of $x$, $y$, $z$, and the second impulse makes up the remaining difference. It is worth noting that the fuel-saving solution favors combined maneuvers, so the first impulse $\Delta v_a$ is written as

$$\Delta v_a = \sqrt{(s_x x)^2 + (s_y y)^2 + (s_z z)^2} \quad (9)$$

Note that there are no constraints on the ranges of $s_x$, $s_y$, $s_z$, which means that the semi-major axis and inclination changes may be larger than the original difference, in order to take advantage of the J2 effect to reduce RAAN differences. This case generally occurs when the required RAAN change is too large, as the allowed transfer time is not sufficiently long.

The control on semi-major axis and inclination by the first impulse leads to a change of RAAN difference during the transfer time $t$ as follows

$$\Delta x = m s_y y + n s_z z \quad (10)$$

where $m = (7\dot{\Omega}_0) \sin i_0 t$, $n = (\dot{\Omega}_0 \tan i_0) \sin i_0 t$ are obtained using Eqs. (4), (7), (8), and $\dot{\Omega}_0$ is the average RAAN rate of the chaser and the target. The new RAAN rate results in a smaller final RAAN difference and lower transfer cost.

The second impulse $\Delta v_b$ can be written as

$$\Delta v_b = \sqrt{(x - s_x x - \Delta x)^2 + (y + s_y y)^2 + (z + s_z z)^2} \quad (11)$$

and the total $\Delta v = \Delta v_a + \Delta v_b$ is

$$\Delta v_a + \Delta v_b = \sqrt{(s_x x)^2 + (s_y y)^2 + (s_z z)^2} + \sqrt{(x - s_x x - \Delta x)^2 + (y + s_y y)^2 + (z + s_z z)^2} \qquad (12)$$

It is difficult to find the minimum of $\Delta v = \Delta v_a + \Delta v_b$ in closed form. Through revisiting to classical Minimum-Inclination Maneuvers by Vallado [12], an analytic approximation becomes available by squaring the two velocities to remove the square roots:

$$\Delta v_a^2 + \Delta v_b^2 = (s_x x)^2 + (s_y y)^2 + (s_z z)^2 + (x - s_x x - \Delta x)^2 + (y + s_y y)^2 + (z + s_z z)^2 \qquad (13)$$

Notice that this approximation ignores the cross product terms $(2\Delta v_a \Delta v_b)$. This allows one to differentiate this expression with respect to $s_x$, $s_y$, $s_z$ and set the derivatives to zero to find the minimum value:

$$\frac{\partial (\Delta v_a^2 + \Delta v_b^2)}{\partial s_x} = 4x^2 s_x + 2mxy s_y + 2nxz s_z - 2x^2 = 0 \qquad (14)$$

$$\frac{\partial (\Delta v_a^2 + \Delta v_b^2)}{\partial s_2} = 2mxy s_x + (4y^2 + 2m^2 y^2) s_y + 2mnyz s_z + 2y^2 - 2mxy = 0 \qquad (15)$$

$$\frac{\partial (\Delta v_a^2 + \Delta v_b^2)}{\partial s_3} = 2nxz s_x + 2mnyz s_y + (4z^2 + 2n^2 z^2) s_z + 2z^2 - 2nxz = 0 \qquad (16)$$

The unknowns $s_x$, $s_y$, $s_z$ can be obtained from Eqs. (14)-(16) as follows

$$s_x = \frac{2x + my + nz}{(4 + m^2 + n^2) x} \qquad (17)$$

$$s_y = \frac{2mx - (4 + n^2) y + mnz}{(8 + 2m^2 + 2n^2) y} \qquad (18)$$

$$s_z = \frac{2nx + mny - (4 + m^2) z}{(8 + 2m^2 + 2n^2) z} \qquad (19)$$

By substituting Eqs. (17)-(19) into Eq. (12), the minimum $\Delta v$ required can be simply approximated. It can be observed that when $s_x$ is zero, both $s_y$, $s_z$ are equal to -1/2. This means that changes of semi-major axis and inclination require two equal values of $\Delta v$. Besides, by using $s_x$, $s_y$, $s_z$ to evaluate $s_x x$ and $x - s_x x - \Delta x$, an interesting identity can be found in the estimated solution, i.e.,

$$s_x x = x - s_x x - \Delta x \qquad (20)$$

which shows that the first and last impulse always provide the same change of RAAN. Equation (20) can be inserted into Eq. (11) to further simplify the estimation expression.

For small eccentricity changes, the required velocity change is given by

$$\Delta v_e = \frac{1}{2} v_0 \sqrt{\Delta e_y^2 + \Delta e_x^2} \qquad (21)$$

where the non-singular equinoctial elements are used for eccentricity, i.e., $e_y = e\sin\omega$, $e_x = e\cos\omega$. An empirical relation, which assumes that the $\Delta v$ for the change of eccentricity vector is divided equally between each impulse, is introduced to account for the additional cost of eccentricity change:

$$\Delta v' = \sqrt{\Delta v_a^2 + (0.5\Delta v_e)^2} + \sqrt{\Delta v_b^2 + (0.5\Delta v_e)^2} \qquad (22)$$

Equation (22) provides an eccentricity correction on the previous equation $\Delta v = \Delta v_a + \Delta v_b$.

## IV. Results

The ninth edition of the global trajectory optimization competition GTCO9 [13] proposed a problem that concerned the removal of 123 debris. A complete summary of the winner solution of the GTOC9 problem by JPL can be found in Ref. [9]. Table 1 and 2 show the transfer durations and $\Delta v$ cost for each mission. Due to the stringent time constraints of the proposed problem, the opportunity of performing optimal-time transfers was extremely rare, and limited time solutions were sought. JPL's solution is here used to evaluate the accuracy of the proposed $\Delta v$ estimation method for time-limited transfers.

Estimations and actual values are usually in good agreement (the average error magnitude is 4.37%), as shown in Table 3. A synoptic view of the missions is provided in Fig. 1, which compares actual and estimated rendezvous $\Delta v$. When taking the additional eccentricity correction into account, the estimated solution becomes even closer to the exact solution with an average error magnitude of 2.83%, as shown in Fig. 2 and Table 4. However, in most cases the estimated solution is slightly below the corresponding exact one. It is worth noting that phasing constraints, which may affect the legs of the optimal GTOC9 solution, are not considered in the approximate analysis. Besides, estimation performance for all the 113 transfer legs (where 123 debris and 10 launches are involved) can also be measured by the mean absolute error (MAE), which is 16.5 m/s and 13.3 m/s for cases without and with eccentricity correction, respectively.

As described in JPL's paper, their method was very good at making large numbers of significant changes to existing solutions, but it had difficulty in finding truly global optima probably due to the existence of multiple local optima in this problem. Therefore, the exact solutions given in Table 2 may not guarantee the minimum, and it could be expected that the exact solution may have room to be improved to some extent.

**Table 1. Transfer Duration Characteristics of JPL's solution**

| Mission | Number of legs | Transfer Duration, days |
|---|---|---|
| 1 | 13 | 24.86,24.98,22.42,24.99,0.29,10.63,25.00,2.70,1.51,1.41,24.67,24.31,5.86 |
| 2 | 11 | 24.93,0.28,0.73,0.39,17.07,1.61,22.42,2.39,15.88,24.97,2.49 |
| 3 | 20 | 14.16,24.94,2.87,8.10,9.00,23.13,23.09,23.09,22.83,24.98,24.98, 24.93,24.94,9.10,13.44,24.99,24.94,24.99,24.98,24.96 |
| 4 | 10 | 23.96,6.48,16.72,23.97,23.95,23.95,23.96,23.99,23.94,23.96 |
| 5 | 13 | 0.45,3.17,24.93,10.34,12.53,7.11,13.44,24.94,24.94,24.98,22.19,24.99,22.01 |
| 6 | 9 | 24.91,0.30,18.39,3.08,20.24,24.96,24.85,24.97,0.28 |
| 7 | 9 | 15.69,0.50,9.83,24.94,24.90,24.48,20.87,24.91,0.66 |
| 8 | 8 | 10.03,24.00,2.83,24.99,24.99,24.96,21.19,24.98 |
| 9 | 11 | 22.69,4.24,24.47,24.46,24.47,24.44,24.46,24.46,24.46,18.54,9.22 |
| 10 | 9 | 0.81,11.59,7.66,1.11,17.46,6.47,20.47,24.47,3.99 |

**Table 2. Mission ΔV Characteristics of JPL's solution**

| Mission | Number of legs | ΔV, m/s | Total ΔV, m/s |
|---|---|---|---|
| 1 | 13 | 161.8,139.2,65.8,208.2,115.2,300.1,564.9,78.3,105.0,233.3,453.5,340.4,300.8 | 3066.5 |
| 2 | 11 | 659.0,301.1,252.1,143.8,146.8,68.6,40.6,84.2,105.3,448.5,148.0 | 2398.0 |
| 3 | 20 | 219.1,80.8,105.2,55.2,140.2,85.5,95.0,237.6,205.9,149.9,245.2,71.6,197.3, 160.4,132.2,240.0,161.2,364.3,230.4,232.5 | 3409.5 |
| 4 | 10 | 86.1,103.1,62.6,222.9,709.1,553.9,219.9,233.9,739.0,232.6 | 3163.1 |
| 5 | 13 | 129.6,45.2,172.9,52.6,160.7,280.8,221.1,163.5,98.2,115.7,164.8,674.8,291.1 | 2571.0 |
| 6 | 9 | 156.0,198.0,305.8,71.2,194.4,920.5,314.1,353.0,272.8 | 2785.8 |
| 7 | 9 | 400.6,173.6,211.3,374.4,109.6,171.2,145.1,194.3,233.0 | 2013.1 |
| 8 | 8 | 287.9,111.9,112.2,144.5,540.0,260.1,198.8,82.7 | 1738.1 |
| 9 | 11 | 83.3,148.1,495.9,464.9,405.2,285.9,254.8,62.3,156.6,36.5,174.9 | 2568.4 |
| 10 | 9 | 189.4,112.9,110.0,121.3,117.9,280.1,300.4,120.6,70.2 | 1422.8 |

**Table 3. Mission ΔV by the Simple Approximation**

| Mission | ΔV, m/s | Total ΔV, m/s | Error,% |
|---|---|---|---|
| 1 | 165.7,140.7,31.49,209.2,109.9,295,562.9,61.85,101.4,226.8,515.7,385.3,298.9 | 3104.9 | **1.25** |
| 2 | 596.5,300.8,249.1,142.9,138.6,66.43,43.64,91.4,108.8,422.2,153.4 | 2313.6 | **-3.52** |
| 3 | 198.7,65.82,96.87,44.67,137.4,53.34,92.44,248.6,204.2,151.2,194.6, 23.93,203.9,166.5,128.6,231.6,160.3,378.8,243,256.9 | 3281.3 | **-3.76** |
| 4 | 89.73,61.54,62.77,230.8,651.5,498.6,203.9,229.2,671.8,224.1 | 2924.0 | **-7.56** |
| 5 | 133.2,37.62,145.8,10.03,185.5,261.7,204.5,108.3,44.47,99.21,165.1,620.5,279.9 | 2296.0 | **-10.7** |
| 6 | 153.1,160.9,313,56.61,204,841.4,304.2,339.8,261.2 | 2634.2 | **-5.44** |
| 7 | 425.8,172.9,218.9,391.3,119.1,174.8,181.8,202.8,214.0 | 2009.3 | **-0.19** |
| 8 | 290.7,117,96.53,144.3,502,251.7,202.7,37.14 | 1642.1 | **-5.52** |
| 9 | 86.24,142.9,458.3,458.3,378.6,312.6,265.2,27.18,162.2,36.86,174.1 | 2502.5 | **-2.56** |
| 10 | 189.0,107.9,94.09,75.6,119.6,287.3,310.3,124.2,69.21 | 1377.2 | **-3.20** |

**Table 4. Mission ΔV by the Simple Approximation with Eccentricity Correction**

| Mission | ΔV, m/s | Total ΔV, m/s | Error,% |
|---|---|---|---|
| 1 | 170.5,145.1,63.13,212.2,115.4,299,564.8,64.33,102.8,229.4,518.4,385.3,300.1 | 3170.6 | **3.39** |
| 2 | 597.8,301.1,249.3,149.8,148.2,71.08,43.65,92.25,108.9,422.4,157.8 | 2342.3 | **-2.32** |
| 3 | 211.7,77.72,97.45,50.57,148.7,85.18,106,252.5,208.2,160.7,218.8, 72.93,206.2,178.8,134.1,241.5,182.4,379.8,243,257.4 | 3513.6 | **3.05** |
| 4 | 89.89,88.69,66.46,232.1,652.3,500.3,204.2,230.4,674.7,241.9 | 2980.9 | **-5.76** |
| 5 | 137.0,38.75,162.9,51.25,186.2,270.6,213.8,134,77.32,118.4,166,621.4,283.0 | 2460.5 | **-4.30** |
| 6 | 155.2,179.2,315.8,81.18,213.3,842,308.3,340.1,271.2 | 2706.2 | **-2.86** |
| 7 | 419.4,171.4,229.4,386.3,117.7,169.3,151.3,199.6,219.8 | 2064.3 | **2.54** |
| 8 | 297.4,118.5,113.6,145.7,502.7,254.9,204.8,64.43 | 1701.9 | **-2.08** |
| 9 | 88.35,149.8,459.6,459.1,384.4,321.6,271.7,49.73,173.1,43.15,174.3 | 2574.7 | **0.25** |
| 10 | 189.4,130.6,117.8,76.58,123.5,292.2,311.3,126,80.48 | 1447.9 | **1.76** |

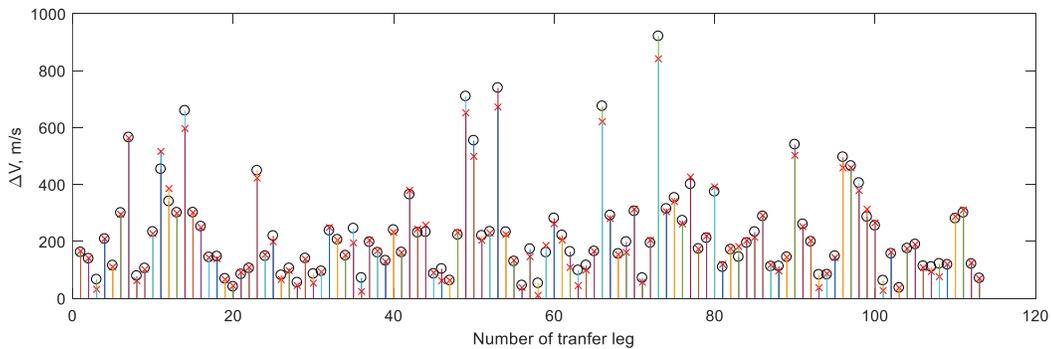

**Fig. 1** Synoptic view of the comparison between exact and estimated solutions (the circle represents JPL's exact solution, and cross is the estimated solution)

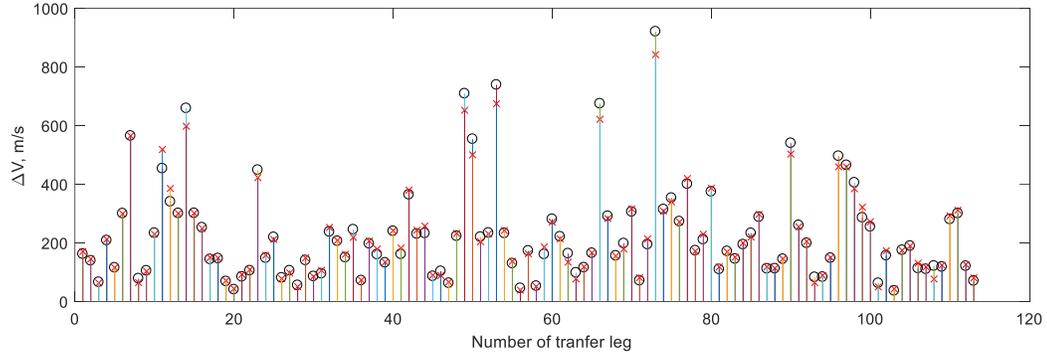

**Fig. 2 Synoptic view of the comparison between exact and estimated solutions (the circle represents JPL's exact solution, and cross is the estimated solution with eccentricity correction)**

Table 5 and Table 6 show two detailed solutions by comparing exact solution and estimation. The transfer time is relatively long (24.86 day) and medium (10.03 day), respectively, for the two solutions. Leg details are not given in Ref. [9], and the exact solutions are here obtained by an optimization procedure involving up to four impulses, by means of a continuous ant colony optimization (ACO) method [5]. The exact solutions by ACO are very close the JPL's exact solutions (which considered up to 5 impulses), though a slightly larger total $\Delta v$ is obtained in both cases. In the exact optimization, multiple impulses are favored in order to match phase and eccentricity. However, the four impulses can be seen as the separation of two single impulses, which are placed at the start and final time of each leg.

In order to obtain a plane alignment, there is a change of RAAN in addition to inclination. This kind of maneuvers do not occur in cases where sufficient transfer time is allowed, and demonstrate that a time-limited solution is required. In the estimated solution, RAAN changes are the same between the initial maneuver and the last maneuver, as illustrated in the previous Section. It is worth noting that, for the transfer described in Table 5, the two-impulse approximation uses a different strategy in terms of semi-major axis change with a small initial increase, but is nonetheless capable of obtaining a similar overall $\Delta v$ cost. In both approximate solutions the semi-major axis is initially increased and then decreased, with the purpose of obtaining the best tradeoff in terms of total $\Delta v$ between changes of RAAN and other parameters (semi-major axis and/or inclination). In other cases, a similar strategy is used for inclination. This fact shows that the semi-major axis and/or inclination changes are not monotonic, which is peculiar in this kind of J2-perturbed time-limited body-to-body transfers.

The root-sum-square (RSR) of individual $\Delta v$ required for the changes of the semi-major axis, RAAN, and inclination, are in good agreement with the corresponding exact $\Delta v_i$. This fact confirms that the $\Delta v$ estimation method, which takes root-sum-square into account for the two impulses, as described in Eqs. (9) and (11), works properly.

Table 5. Detailed solution of the first leg of mission-1 (debris 23 to 55)

| Solution | | time, day | $\Delta v_i$, m/s | $\Delta a$, km | $\Delta i$, deg | $\Delta \Omega$, deg | RSR $\Delta v$, m/s | Total $\Delta v$, m/s |
|---|---|---|---|---|---|---|---|---|
| Exact | 1 | 0.275 | 134.2 | -37.08 | -0.999 | -0.083 | 132.2 | 164.55 |
| | 2 | 0.312 | 0.818 | 1.196 | -0.0037 | -0.0014 | 0.811 | |
| | 3 | 24.828 | 15.41 | 8.671 | -0.114 | 0.005 | 15.56 | |
| | 4 | 24.860 | 14.12 | -18.23 | -0.076 | 0.023 | 14.08 | |
| Estimated | 1 | 0 | 98.17 | 13.59 | -0.743 | -0.104 | 98.17 | 165.73 |
| | 2 | 24.86 | 67.56 | -59.03 | -0.449 | -0.104 | 67.56 | |

Table 6. Detailed solution of the first leg of mission-8 (debris 86 to 34)

| Solution | | time, day | $\Delta v_i$, m/s | $\Delta a$, km | $\Delta i$, deg | $\Delta \Omega$, deg | RSR $\Delta v$, m/s | Total $\Delta v$, m/s |
|---|---|---|---|---|---|---|---|---|
| Exact | 1 | 0.0667 | 141.30 | 54.82 | -0.778 | -0.728 | 141.26 | 294.06 |
| | 2 | 0.102 | 76.75 | -39.67 | -0.432 | -0.383 | 77.80 | |
| | 3 | 9.997 | 33.32 | -44.15 | -0.00027 | -0.187 | 33.45 | |
| | 4 | 10.03 | 42.70 | -57.10 | -0.0190 | -0.233 | 42.52 | |
| Estimated | 1 | 0 | 165.00 | 70.71 | -1.016 | -0.705 | 165.00 | 290.73 |
| | 2 | 10.03 | 125.73 | -156.82 | -0.213 | -0.705 | 125.73 | |

## V. Conclusions

A procedure has been developed for preliminary evaluation of transfer costs between objects in LEO, in order to speed up the analysis of debris removal missions aiming at multiple objects, when they must be selected in a large set of targets with similar orbit inclination. The analysis tool that has been developed is in particular beneficial in the preliminary design phases, when mission feasibility must be assessed and trade-offs are required, and a precise evaluation of the costs can reduce design uncertainties. The good accuracy of the proposed estimation method has been verified by comparison with precisely optimized solutions. Even though the estimation is in general slightly below the exact solution, the benefits, which can be obtained when a proper analytical approximation of transfer cost is used in multiple-target active debris removal, are thus demonstrated.